\documentclass[aps,prl,twocolumn,groupedaddress,showpacs]{revtex4}

\usepackage{graphicx}

\usepackage{mathrsfs}

\begin{document}


\title{Truly Random Number Generation Based on Measurement of Phase Noise of Laser}


\author{Hong Guo}
\email[Corresponding author:\ ]{hongguo@pku.edu.cn}
\author{Wenzhuo Tang}
\author{Yu Liu}
\author{Wei Wei}

\affiliation{CREAM Group, State Key Laboratory of Advanced Optical
Communication Systems and Networks (Peking University) and Institute
of Quantum Electronics, School of Electronics Engineering and
Computer Science, Peking University, Beijing 100871, China}


\date{\today}

\begin{abstract}We present a simple approach to realize truly random number generation
based on measurement of the phase noise of a single mode vertical
cavity surface emitting laser (VCSEL). The true randomness of the
quantum phase noise originates from the spontaneous emission of
photons and the random bit generation rate is ultimately limited
only by the laser linewidth. With the final bit generation rate of
20 Mbit/s, the physically guaranteed truly random bit sequence
passes the three standard random tests. Moreover, for the first
time, a {\it continuously} generated random bit sequence up to 14
Gbit is verified by two additional criteria for its true randomness.
\end{abstract}




\maketitle


Random number generator (RNG) has wide applications in statistical
sampling \cite{L99}, computer simulations \cite{G03}, randomized
algorithm \cite{M05} and cryptography \cite{M97}. Traditionally,
pseudorandom number generator (PRNG) based on computational
algorithms is adopted to generate random bits and is competent in
many fields. However, it cannot produce truly random (unpredictable
and irreproducible) bit sequence, and so may result in potential
dangers in security related applications, say, in quantum
cryptography \cite{B92}. Actually, the unconditional security of
quantum key distribution can ONLY be guaranteed when a truly random
number generator (TRNG), based on quantum mechanical process instead
of the intractability assumption with classical algorithms
\cite{B86}, is available.

Distinct from PRNG, a TRNG can only be realized by a physical way,
instead of an algorithm-based way; however, a physical way does not
sufficiently guarantee the true randomness. The physically random
processes, such as radioactive decay \cite{S70}, electric noise in
circuits \cite{P00}, frequency jitter of electric oscillator
\cite{B03}, and those based on laser (photon) emission/detection
\cite{Sti07,D08,W09}, can ensure the inability of pre-estimation on
random numbers and so can be adopted as candidates to implement
TRNG. In particular, those based on the detection of laser field
attracted tremendous interests in recent decade. Recently, chaotic
laser, with ultra-wide bandwidth, becomes a promising candidate for
GHz random bit generation \cite{U08,R09}. However, since the signal
of chaotic laser has a periodicity originated from the photon round
trip time, it is essentially NOT a truly random source. Also, we
know, the chaotic systems are deterministic that looks random but
without inherent randomness \cite{CW09,J00}. Hence, we coin this
kind of physically-based, rather than algorithm-based, pseudo RNG as
\emph{physically pseudorandom number generator (PPRNG)}. Thus, the
true randomness guaranteed by physical principle(s), instead of its
generation rate, should be firstly pursued for a TRNG, otherwise,
even though with ultrahigh rate, it is just a pseudo RNG; on the
other hand, the TRNGs based on the above-mentioned physical
mechanisms \cite{S70,P00,B03,Sti07,D08,W09} cannot offer the bit
generation rate as high as PPRNG based on chaotic lasers
\cite{U08,R09}. So far, the typical maximal random bit generation
rate is around 10 Mbit/s for electric oscillator jitter measurement
scheme \cite{B03} and 4 Mbit/s for photon detection scheme
\cite{D08}. Also, it should be noted that in these schemes, the
statistical bias and correlation for long random bit sequence were
not investigated.

  \begin{figure}
  \centering
  \includegraphics[width=3.5in]{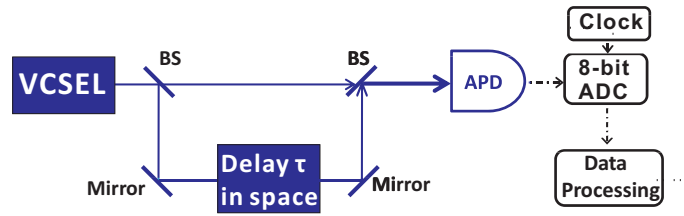}
  \caption{(color online). Schematic setup of TRNG based on the phase noise measurement
  using delayed self-homodyne method. BS, beam splitter; APD, avalanche photodetector with the
  low (high) cutoff frequency of 50 kHz (1 GHz). ADC, 8-bit binary
  analog-digital-converter working at 40 MHz.\label{fig1}}
  \end{figure}

In this Letter, we propose a new and simple TRNG scheme based on the
true randomness of the quantum phase noise, which is a Gaussian
random variable \cite{L67,H82}, of a single-mode vertical cavity
surface emitting laser (VCSEL). The true randomness of the quantum
phase noise is originated from the random nature of spontaneous
emission and is guaranteed by physical principle. It will, in the
following, be shown that the random bit generation rate of this TRNG
is ultimately limited only by the laser linewidth. In our
experiment, the high final bit generation rate reaches 20 Mbit/s
with guaranteed true randomness. Further, the true randomness is not
only guaranteed in physical principle and standard tests, but is
also verified by two additional criteria (statistical bias and
correlation) for the long random bit sequence up to 14 Gbit, for the
first time.

 \begin{figure}
  \includegraphics[width=3.3in]{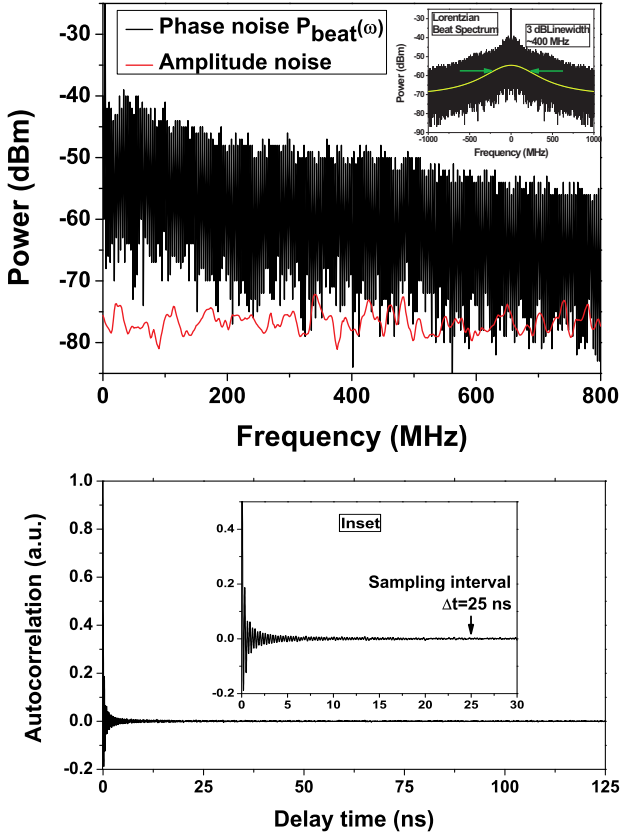}
  \caption{(color online). (a) The quantum phase (classical amplitude) noise of the laser
  field is observed with (without) the beat signal. The inset is the power spectral density of the beat signal.
  (b) Autocorrelation function of the beat signal versus time interval. In our experiment, the sampling
  interval of 25 ns (40 MHz sampling rate) is chosen.}  \label{fig2}
  \end{figure}


The schematic setup is shown in Fig. \ref{fig1} and the delayed
self-homodyne method is used to measure the phase noise of the
VCSEL. In this case, the output alternative current (AC) voltage of
the avalanche photodetector (APD) detecting the beat signal is
$V_{ph} \propto {\rm
AC}[I_{beat}]=2\mathscr{E}(t)\mathscr{E}(t+\tau)\cos[\phi(t)-\phi(t+\tau)],$
where the amplitude fluctuations of $\mathscr{E}(t)$ and
$\mathscr{E}(t+\tau)$ are negligible compared to the phase
fluctuation corresponding to $\cos[\phi(t)-\phi(t+\tau)]$
\cite{L67,H82}. When the delay time is much longer than the
coherence time of laser (i.e., $\tau\gg\tau_{coh}$), the phase
difference $\Delta \phi(t) = \phi(t) - \phi (t + \tau)$ is a
Gaussian random variable, and thus the autocorrelation function of
the electric field of the laser is eliminated \cite{L67}, i.e.,
\begin{eqnarray}\label{eq1}
\langle E^*(t) E(t+\tau)\rangle \propto \exp(-|\tau|/\tau_{coh})\rightarrow
0,
\end{eqnarray}
where $\tau_{coh} = (\pi \Delta \nu_{laser})^{-1}$ \cite{H82}, and
$\Delta \nu_{laser}$ is the laser linewidth. This indicates that the
electric field amplitudes of the laser at different time are
mutually independent, if time interval is much longer than the
coherence time of the laser. Further, similar calculation procedure
can be applied to obtain the autocorrelation function of the beat
signal [$E_{beat}(t)$] as $\langle E_{beat}^*(t) E_{beat}(t+\Delta
t)\rangle$, where $\Delta t$ is the sampling interval for original
random bit generation. Using $E_{beat}(t) = E(t) + E(t+\tau)$ and
Eq. (\ref{eq1}), it is evident that when the sampling time $\Delta
t$ meets $\Delta t \gg \tau + \tau_{coh}$, the autocorrelation of
the beat signal will also be eliminated. Thus, the bits extracted
from the beat signal are mutually independent and can be adopted to
implement TRNG.

 \begin{figure}
  \includegraphics[width=3.3in]{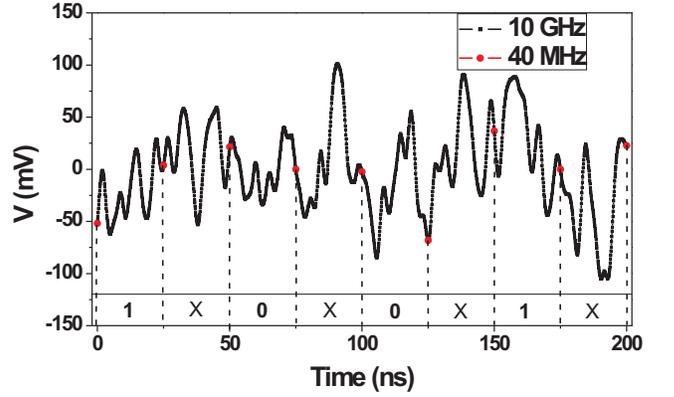}
  \caption{(color online). A 200 ns trace of the APD-detected voltages of the beat signal (small black dots) is
  recorded at 10 GHz, while the random signal (big red dots) is sampled at 40 MHz rate (25 ns interval). The
  final random bit is obtained from the least significant bit (LSB, i.e., its parity) of a sequence of 8-bit
  binary derivatives obtained by performing subtraction between two consecutive sampled voltages (shown in
  the bottom strip).}
  \label{fig3}
  \end{figure}


%
%

 \begin{figure}[htbp]
  \includegraphics[width=3.3in]{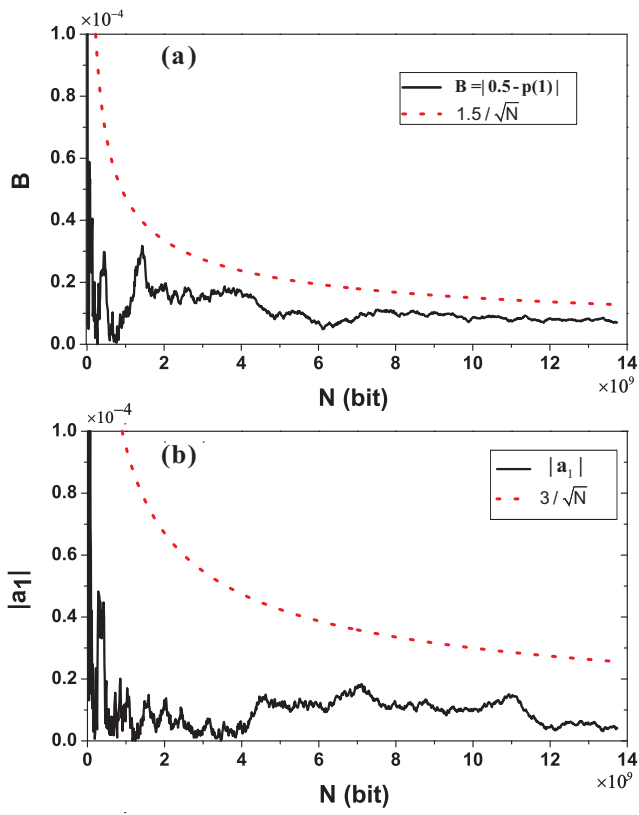}
  \caption{(color online). (a) The statistical bias ($B$) of the final random bit sequence. It can be seen that $B<1.5/\sqrt{N}$
  always holds and converges to zero for large bit sequence, where $p(1)$ is the probability of ones
  in sequence. (b) The absolute value of the first-order
  correlation coefficient $|a_1|$ of the final random bit sequence. It can be seen that $|a_1|<3/\sqrt{N}$ always holds and $|a_1|$
  converges to zero for large bit sequence.}
  \label{fig5}
  \end{figure}

In experiment (Fig. \ref{fig1}), a $795\ \rm{nm}$ VCSEL laser works
at 1.5 mA, a little above the threshold current 1.0 mA. The laser
linewidth $\Delta \nu_{laser} = 200\ \rm{MHz}$ ($\tau_{coh} = 1.59\
\rm{ns}$) of laser is inversely proportional to the laser power,
while the classical noises are independent on it \cite{QI}. Due to
working just above threshold, the {\it quantum} phase noise of laser
dominates over the {\it classical} amplitude noise to ensure the
true randomness of generated bits. The delay time $\tau$ is set to
be about $10\ \rm{ns}$ (corresponds to 3.0 m space delay) in order
to fulfill $\tau\gg{\tau_{coh}}$. So, the self-homodyne method with
delay time $\tau$ is used to obtain the beat signal with 3 dB
linewidth of $400\ \rm{MHz}$ (detected by an APD) and its power
spectral density is shown in the inset of Fig. \ref{fig2} (a). It
can be seen, from Fig. \ref{fig2} (a), that the {\it classical}
amplitude fluctuation is negligible compared to the {\it quantum}
phase fluctuation within 200 MHz (the gap is about 20 dB). Using
Wiener-Khintchine theorem \cite{W30,K34}, i.e.,
\begin{eqnarray}\label{eq3}
R_{beat} (t)=\int_{-\infty}^{+\infty}P_{beat} (\omega)\exp(-i\omega
t) d\omega,
\end{eqnarray}
the autocorrelation function [$R_{beat}(t)$] of the beat signal
 is obtained from the power spectral density of the phase noise
[$P_{beat}(\omega)$ in Fig. \ref{fig2} (a)] and illustrated in Fig.
\ref{fig2} (b). It can be seen from Fig. \ref{fig2} (b) that the
autocorrelation of the beat signal can be ignored, if the sampling
interval is set as $\Delta t \gg \tau + \tau_{coh}$. In our
experiment, the sampling rate is chosen as $40\ \rm{MHz}$
accordingly, i.e., $\Delta t = 25\ \rm{ns}$, so the bits extracted
from these sampled voltages are mutually independent. These sampled
voltages are digitized by an 8-bit analog-digital-converter (ADC)
shown as the red dots in Fig. \ref{fig3}, and further we have
confirmed that the distribution of voltages is symmetric. Hence, we
take the least significant bit (LSB) of each sampled 8-bit voltage
as the original random bit, i.e., the parity of this 8-bit binary
number, which represents whether the voltage falls in an even or odd
bin of the total 256 bins. For the non-ideal distribution of these
voltages, the probability of all the even and odd bins of the total
256 bins are not perfectly equal, and the bit sequence shows a
statistical bias of the order of $10^{-3}$. For much lower bias, we
perform a subtraction between two consecutive sampled voltages to
obtain a sequence of ${N}/{2}$ 8-bit binary derivatives as $V_2-V_1,
V_4-V_3, \ldots, V_N-V_{N-1}$, where $N$ is the total number of the
original sampled voltages. In this process, each voltage is used
only once, and thus no correlation is introduced. After that, we
adopt the LSB of the 8-bit binary derivatives to generate the final
random bits. Therefore, we directly obtained the final random bit at
generation rate of 20 Mbit/s with a software-based post-processing.
Note that, the post-processing enhances the performance of the
random bits sequence by lowering the statistical bias while not
introducing any additional correlations. For a TRNG, both the
statistical bias and the absolute value of the first-order
correlation coefficient of the final random bit sequence are
expected to be smaller than three standard deviations
($3\sigma_1=1.5/\sqrt{N}$ for statistical bias [Fig. \ref{fig5}(a)],
and $3\sigma_2=3/\sqrt{N}$ for correlation coefficient [Fig.
\ref{fig5}(b)]) with the probability of 99.7\%. In our case, both
criteria are well satisfied for the final random bit sequence up to
14 Gbit.

We continuously record a final random bit sequence of 1 Gbit, which
passed three standard random tests, i.e., ENT \cite{ENT}, Diehard
\cite{Diehard} and STS \cite{STS}. The ENT results are: Entropy $=
1.000000$ bit per bit (the optimum compression would reduce the bit
file by $0\%$). $\chi^2$ distribution is $0.53$ (randomly would
exceed this value by $46.62\%$ of the times). Arithmetic mean value
of data bits is $0.5000$. Monte Carlo value for $\pi$ is
$3.141725650$. Serial correlation coefficient is $-0.000017$. The
Diehard and STS test results are shown in Tables I and II,
respectively.


%

\begin{table}
  \centering
  \caption{Results of Diehard statistical test suite. Data sample containing 100 Mbits
  is used for the Diehard test. For the cases of multiple $p$-values, a
  Kolmogorov-Smirnov (KS) test is used to obtain a final $P$-value, which measures the uniformity of the
multiple $p$-values. The test is considered successful if all final
$P$-values satisfy $0.01\leq P\leq 0.99$.}
\begin{ruledtabular}
\begin{tabular}{lllll}
    Statistical test                 & $P$-value        & Result   \\
    \hline
    Birthday spacings                &0.910531 [KS]  & Success \\
    Overlapping permutations         &0.294899      & Success  \\
    Ranks of $31\times31$ matrices   &0.322213      &Success   \\
    Ranks of $32\times32$ matrices   &0.482575        &Success    \\
    Ranks of $6\times8$ matrices     &0.749427 [KS]   &Success   \\
    Monkey tests on 20-bit words     &0.019887 [KS]   &Success  \\
    Monkey test OPSO                 &0.079864 [KS]  &Success \\
    Monkey test OQSO                 &0.725649 [KS]   &Success \\
    Monkey test DNA                  &0.293543 [KS]   &Success\\
    Count 1's in stream of bytes     &0.244463         &Success\\
    Count 1's in specific bytes      &0.062188 [KS]     &Success\\
    Parking lot test                 &0.806898 [KS]   & Success\\
    Minimum distance test            &0.326209 [KS]   &Success \\
    Random spheres test              &0.902946 [KS] &  Success \\
    Squeeze test                     &0.815876 [KS]   &Success  \\
    Overlapping sums test            &0.806025 [KS]   &Success    \\
    Runs test (up)                   &0.817356        &Success \\
    Runs test (down)                 &0.805323        &Success \\
    Craps test No. of wins           &0.502035        &Success \\
    Craps test throws/game           &0.403322        &Success \\
\end{tabular}
\end{ruledtabular}
\end{table}

\begin{table}[htbp]
  \centering
  \caption{Results of NIST statistical test suite. Using 1000 samples
  of 1 Mbits data and significance level $a=0.01$, for ``Success", the
  $P$-value (uniformity of $p$-values) should be larger than 0.0001 and
  the proportion should be greater than 0.9805608 \cite{STS}. For the tests
  which produce multiple $P$-values and proportions, the worst case is
  shown. As advised by NIST, the Fast Fourier Transform test is
  disregarded \cite{Claim}.}
\begin{ruledtabular}
  \begin{tabular}{lllll}
    Statistical test            &$P$-value         &Proportion    &Result  \\ \hline
    Frequency                   & 0.679846         &0.9916  & Success   \\
    Block frequency             & 0.248571             &0.9897   & Success \\
    Cumulative sums             & 0.858032       &0.9888  & Success  \\
    Runs                        & 0.816029       &0.9907  &Success   \\
    Longest run                 &  0.648795      &0.9935   &Success    \\
    Rank                        & 0.609895     &0.9860  &Success   \\
    Nonperiodic       &  0.569334      &0.9823 &Success  \\
    Overlapping       &0.565500    &0.9916  &Success \\
    Universal                     &0.143336    &0.9888 &Success \\
    Approximate          &0.590520    &0.9879  &Success\\
    Random excursions              &0.016388         &0.9880  &Success\\
    Random variant      &0.029796     &0.9865   &Success\\
    Serial                         &0.946683    &0.9916  & Success\\
    Linear complexity              &0.732979    &0.9915  &Success \\
 \end{tabular}
\end{ruledtabular}
\end{table}


It should be noted that, for a nonuniform distribution of the
probability of 256 8-bit binary derivatives, if more than 1 bit are
extracted from each 8-bit binary derivatives in order to improve the
random bit generation rate (see, e.g., 5 LSBs are adopted in
\cite{R09}), an additive correlation in the final random bit
sequence will be introduced, even though this additive correlation
is not so significant to fail the random tests. Taking 5 LSBs for an
instance, every set of the 5 LSBs possesses a different probability
(due to the nonuniform distribution) and thus these 5 bits from the
same set are correlated to some extent. However, with this additive
correlation within the same set, both the random bit sequence of
extracting 5 LSBs (with the sampling rate of 2.5 GHz in \cite{R09})
and 4 LSBs (with the sampling rate of 40 MHz in our case) from an
8-bit binary number both successfully pass the three standard random
tests. This fact also indicates that the standard random tests are
only a way to examine whether the random bit stream is
``sufficiently" random, but not to judge whether it is truly random.

We propose a new and simple approach to realize a high-speed TRNG,
which is compact and convenient to implement. The randomness of our
TRNG is physically guaranteed by the {\it intrinsic} random nature
of the quantum phase noise originated from the spontaneous emission
of photons. Moreover, for the first time, the true randomness is
verified by both the statistical bias and the correlation
coefficient for long random bit sequence up to 14 Gbit. Note that,
the long random bit sequence is even more important than generation
rate, because it is the length of the random bit sequence that is
required in most applications and essentially, it is a metric for
qualifying the true randomness. Compared to the chaotic laser, the
intrinsic phase noise of a free-running laser is confirmed in true
randomness, which only depends on its inherent quantum mechanical
properties and does not need the external optical feedback to laser
thereby introducing photon round trip period. Although the random
bit generation rate is not as high as that of chaotic laser scheme
\cite{U08,R09}, its physically guaranteed true randomness and high
generation rate, together with its simplicity and compactness, are
attractive for applications which need true randomness. Also, a
higher generation rate is attainable using a laser with larger
linewidth and faster data acquisition hardware.

This work is supported by the Key Project of National Natural
Science Foundation of China (NSFC) (grant 60837004). We acknowledge
the support from W. Jiang, K. Deng, X. X. Liu, C. Zhou, and
G. D. Xie, and the helpful discussions with B. Luo and J. B. Chen.

\end{document}